\documentclass[12pt]{article}
\usepackage[margin=1in]{geometry}
\usepackage{amsmath,amssymb,amsfonts}

\usepackage{times}

\newcommand{\DVARS}{\ensuremath{{\mbox{DVARS}}}}

\newcommand{\Var}{\ensuremath{\mathsf{Var}}}
\newcommand{\Expd}{\ensuremath{\mathsf{E}}}
\newcommand{\Corr}{\ensuremath{\mathsf{Corr}}}
\newcommand{\Cov}{\ensuremath{\mathsf{Cov}}}

\begin{document} 

\title{\bf Notes on Creating a Standardized Version of DVARS}
\author{Thomas Nichols\thanks{Warwick Manufacturing Group and Department of Statistics,  University of Warwick, Coventry, CV4 7AL, U.K.}
}
\maketitle

\begin{abstract} 
By constructing a sampling distribution for DVARS we can create a
standardized version of DVARS that should be more similar across
scanners and datasets.
\end{abstract}



\noindent Please note: The content of this document is identical to the the version dated September 12, 2013, available at
  http://warwick.ac.uk/tenichols/scripts/fsl/StandardizedDVARS.pdf. This
  document is being posted to arXiv to make it more accessible, in
  advance of further work on DVARS entited ``Insight and Inference for DVARS''
  by Afyouni \& Nichols.

\section{Introduction}

Power et al. (2012) proposed a measure to characterize the
quality of fMRI data, an image-wide summary that produces
a time series that can detect scans that are corrupted by artifacts.
They called their measure DVARS, the per-image standard deviation of
the temporal derivative of the data.  
Since at least 2006 Matthew Brett's
Data Diagnostics webpage\footnote{{\tt
    http://imaging.mrc-cbu.cam.ac.uk/imaging/DataDiagnostics}, viewed
  28 October, 2012; it lists the ``last edited'' data as 31 July
  2006.  See also a new implementation in the spmttools toolbox, {\tt http://sourceforge.net/projects/spmtools}} 
has offered tsdiffana.m, a Matlab script that produces the same
measure.  For simplicity, I'll stick with the snappier name DVARS.

While DVARS does an excellent job of detecting bad scans--bad pairs of
scans actually--it does not have any absolute units.  The average
value of DVARS (on good data) will depend on the temporal standard
deviation and the temporal autocorrelation of the data.  The purpose
of this short note is to describe a formal description of DVARS, its
nominal standard deviation, which leads to a standardized version of
DVARS which should be more comparable and interpretable between sites
and scanners.  

\section{Methods}

Let $Y_{i,t}$ be fMRI time series data, for voxels $i=1,...,I$
and time points $t=1,...,T$.  A reasonable, approximate model for
well-behaved (outlier-free data) is 
\begin{equation}
  Y_{i,t} = \mu_i + \epsilon_{i,t}
\end{equation}
where $\mu_i$ is the constant, T2* image of the brain and
$\epsilon_{i,t}$ is the  noise.  Of course this model is
wrong, as it neglects the experimental variation, but we expect such
effects to be trivial relative to the artefactual variation of
interest.  The issue of drift and temporal
autocorrelation will be addressed shortly.

DVARS is based on the spatial standard deviation of the temporal difference image\footnote{
For simplicity this sample variance doesn't include the term where the
mean of the difference is subtracted out, since
$\Expd(Y_{i,t}-Y_{i,t-1})$ should be zero by our model.}:
\begin{eqnarray}
  \DVARS_t  &=& \sqrt{\frac{1}{I}\sum_i\left( Y_{i,t}-Y_{i,t-1} \right)^2}. \\
            &=& \sqrt{\frac{1}{I}\sum_i\left( \epsilon_{i,t}-\epsilon_{i,t-1} \right)^2}.
\end{eqnarray}
That is, the magic of DAVARS is that the differencing cancels out $\mu_i$,
the T2* brain.

The problem of predicting a null, default behavior of DVARS, however,
is that it depends on the variance of spatial noise, and the spatial
noise structure is complicated and hard to model in general.
In the time domain, however, we have a reasonable working model of the noise, the
Auto Regressive order-1 (AR(1)) model.  If we can assume that
the spatial and temporal noise structure doesn't
interact\footnote{Formally, ``doesn't interact'' means that the
spatiotemporal correlation structure is separable into a product of
spatial and temporal components.  Again, for well-behaved data this is
a reasonable assumption.}, then standardizing the noise variance in
the time domain will result in unit variance in the spatial domain.

First, we need to state the AR(1) model and see how it predicts the variance
of the temporal difference data. If each voxel's noise
follows an AR(1) model then for voxel $i$ we have
\begin{eqnarray}
  \Var(Y_{i,t})&=&\sigma^2_i\\
  \Corr(Y_{i,t},Y_{i,t-1})&=&\rho_i.
\end{eqnarray}
These parameters are easily estimated as the usual standard deviation
and $\rho$ is even available from {\tt fslmaths} with the {\tt -ar1}
option.  
To determine the variance of the
temporal difference we need a basic result
from probability:
For two correlated random variables $A$ \& $B$, the variance of their
difference is the sum of their variances minus twice their covariance:
$$
  \Var(A-B)= \Var(A)+\Var(B)-2\Cov(A,B)
$$

Thus if the time series follow an AR(1) model, the differenced times
series have variance
\begin{eqnarray}
  \Var( Y_{i,t}-Y_{i,t-1}) &=& 2\sigma_i^2 -2\rho_i\sigma_i^2\\
             &=&2(1-\rho_i)\sigma_i^2.
\end{eqnarray}
This means we can predict the expected value of squared $\DVARS$:
\begin{eqnarray}
  \Expd(\DVARS^2_t)  &=& \frac{1}{I}\sum_i\Expd\left( (Y_{i,t}-Y_{i,t-1})^2 \right) \\
            &=& \frac{1}{I}\sum_i\Var\left( Y_{i,t}-Y_{i,t-1} \right)\\
            &=& \frac{1}{I}\sum_i 2(1-\rho_i)\sigma_i^2.
\end{eqnarray}
This leads to the following revised definition of DVARS as 
\begin{eqnarray}
  \DVARS^*_t  &=& \frac{\sqrt{ \frac{1}{I}\sum_i\left( Y_{i,t}-Y_{i,t-1} \right)^2}}{
    \sqrt{ \frac{1}{I}\sum_i 2(1-\rho_i)\sigma_i^2 }
  }.
\end{eqnarray}
This is easily implemented because $\sigma_i$ and $\rho_i$ can be
computed in {\tt fslmaths}.  Further, since we're worried about
outliers, we can make use of robust estimators of $\sigma_i$ and
$\rho_i$.  The simplest robust estimator of standard deviation is
based on the Inter-Quartile Range (IQR), based on the following
relationship for Normal variates:
$$
  \sigma = IQR/1.349
$$
This is what I have implemented in my own DVARS script\footnote{{\tt http://go.warwick.ac.uk/tenichols/scripts/fsl/DVARS.sh}}.  Unfortunately I
haven't found a similar simple robust estimate for the AR(1)
coefficient $\rho_i$, and thus have implemented the standard estimate.

Finally, with knowledge of the variance of the difference at each
voxel $i$, we can also propose a new variant of DVARS based on
voxel-wise standardized difference data:
\begin{eqnarray}
  \DVARS^{**}_t  &=& \sqrt{ \frac{1}{I}\sum_i \left(\frac{ Y_{i,t}-Y_{i,t-1} }{ \sqrt{ 2(1-\rho_i)\sigma_i^2}}\right)^2}.
\end{eqnarray}
However, this may not be as sensitive to problems because it will
down-weight the voxels with high variance, i.e.\ those around the edge
of the brain.  On the other hand, since most of the edge-related
variance is going to be due to motion and we already have the motion
predictors, $\DVARS^{**}_t$ may be more useful for picking up problems
that are {\em not} related to motion.




\section{Discussion}

The principal limitations of this work is that it depends on an
estimates of standard deviation and AR-1 coefficient that are not
themselves corrupted by bad data.  Further work is needed to identify a
robust estimator of $\rho$.  Also, any sensible time series modelling
effort begins by regressing out a linear from the data, and as fMRI is
susceptible to drift, perhaps standard drift modelling should be
done.  As DVARS is driven by the most short-scale changes possible
(from time $t$ to $t+1$) it won't be affected by removal of drift, but
it may result in more accurate modelling of the temporal correlation,
and thus more accurate standardization.

Another limitation is that some users may {\em like} how DVARS
reflects the underlying time series variance $\sigma$, and watch how
the absolute value of DVARS changes with subsequent preprocessing
steps.  My response to this is that it is not just $\sigma$ that
changes but also the correlation, something that users may have less
intuition on.  And further, if the variance (and autocorrelation) are
of interest, they ideally should be separately plotted and recorded, instead
of indirectly inferred through DVARS.

\section*{Acknowledgments}
I am grateful to the following people at Washington University at
St. Louis who gave me feedback on this work:  Jonathan Power, Matt
Glasser, Deanna Bartsch and Steve Peterson. 

\section*{References}

Power, J. D., Barnes, K. A, Snyder, A. Z., Schlaggar, B. L., \& Petersen, S. E. (2012). Spurious but systematic correlations in functional connectivity MRI networks arise from subject motion. NeuroImage, 59(3), 2142-54.



\end{document}